\documentclass[usenatbib]{mnras}

\pdfoutput=1
\usepackage{amssymb,amsmath}
\usepackage{graphicx}
\usepackage{multirow}
\usepackage{natbib}
\usepackage{tabularx}
\usepackage[bottom]{footmisc}
\usepackage{wasysym}
\usepackage{upgreek}
\usepackage{color}

\def\dm{\mathrm{pc}\,\mathrm{cm}^{-3}}

\title[SUPERB III]{The SUrvey for Pulsars and Extragalactic Radio Bursts III: Polarization properties of FRBs 160102 \& 151230}

\author[M.~Caleb et al.]
{M.~Caleb$^{1,2,3,4}$\thanks{Email: manisha.caleb@manchester.ac.uk},
E. F.~Keane$^{5, 3, 4}$, 
W.~van Straten$^{6}$, 
M.~Kramer$^{7,1}$, 
J. P.~Macquart$^{8}$, 
\newauthor M.~Bailes$^{3, 4}$,
E. D.~Barr$^{7,3,4}$,
N. D. R.~Bhat$^{4, 8}$,
S.~Bhandari$^{3,4,9}$,
M.~Burgay$^{10}$,
\newauthor W.~Farah$^{3}$,
A.~Jameson$^{3,4}$,
F.~Jankowski$^{1,3,4}$,
S.~Johnston$^{9}$,
E.~Petroff$^{11,3,4,9}$,
\newauthor A.~Possenti$^{10}$,
B.~Stappers$^{1}$,
C.~Tiburzi$^{12}$,
V.~Venkatraman Krishnan$^{3,4}$
\\ \\
$^{1}$ Jodrell Bank Centre for Astrophysics, School of Physics and Astronomy, The University of Manchester, Manchester M13 9PL, UK \\
$^{2}$ Research School of Astronomy and Astrophysics, Australian National University, ACT, 2611, Australia\\
$^{3}$ Centre for Astrophysics and Supercomputing, Swinburne University of Technology, P.O. Box 218, Hawthorn, VIC 3122, Australia \\
$^{4}$ ARC Centre of Excellence for All-sky Astrophysics (CAASTRO) \\
$^{5}$ SKA Organisation, Jodrell Bank Observatory, SK11 9DL, UK \\ 
$^{6}$ Institute for Radio Astronomy \& Space Research, Auckland University of Technology, Private Bag 92006, Auckland 1142, New Zealand \\
$^{7}$ Max-Planck-Institut f\"ur Radioastronomie, Auf dem H\"ugel 69, D-53121 Bonn, Germany  \\ 
$^{8}$ International Centre for Radio Astronomy Research, Curtin University, Bentley, WA 6102, Australia \\ 
$^{9}$ CSIRO Astronomy \& Space Science, Australia Telescope National Facility, P.O. Box 76, Epping, NSW 1710, Australia \\ 
$^{10}$ INAF-Osservatorio Astronomico di Cagliari, Via della Scienza 5, I-09047 Selargius (CA), Italy \\ 
$^{11}$ ASTRON, The Netherlands Institute for Radio Astronomy, Postbus 2, 7990 AA Dwingeloo, The Netherlands\\
$^{12}$ Fakult\"at fur Physik, Universit\"at Bielefeld, Postfach 100131, D-33501 Bielefeld, Germany }

\begin{document}
\maketitle

\begin{abstract}
We report on the polarization properties of two fast radio bursts (FRBs): 151230 and 160102 discovered in the SUrvey for Pulsars and Extragalactic Radio Bursts (SUPERB)
at the Parkes radio telescope. FRB 151230 is observed to be $6 \pm 11 \%$ circularly polarized and  $35 \pm 13$ \% linearly polarized with a rotation measure (RM) consistent with zero.
Conversely, FRB 160102 is observed to have a circular polarization fraction of $30 \pm 11$ \%, linear polarization fraction of $84 \pm 15$ \% for RM $= -221(6)$ rad m$^{-2}$ and 
the highest measured DM ($2596.1 \pm 0.3$ pc cm$^{-3}$) for an FRB to date. We examine possible progenitor models for FRB 160102 in extragalactic, 
non-cosmological and cosmological scenarios. After accounting for the Galactic foreground contribution, we estimate the intrinsic RM to be $-256(9)$ rad m$^{-2}$ in the
low-redshift case and $\sim -2.4 \times 10^{2}$ rad m$^{-2}$ in the high-redshift case. We assess the relative likeliness of these scenarios and how each can be tested.
We also place constraints on the scattering measure and study the impact of scattering on the signal's polarization position angle.

\end{abstract}


\section{Introduction}

Radio astronomy is exploring --- with progressively increasing effectiveness --- an intriguing phase space 
of the dynamic Universe on timescales of milliseconds. The recent development of sensitive, high
time-resolution instruments over the last decade has enabled the discovery of millisecond 
duration fast radio bursts \citep[FRBs;][]{Lorimer}. FRBs are characterised by 
dispersion measures (DMs) which are much too large to be accounted for by the Galaxy \citep{Thornton, Champion}. All FRBs
exhibit a power law dependent time delay consistent with $\nu^{-2}$, which is characteristic of propagation through a cold ionised diffuse medium.

FRB theories can be broadly classified into two categories: cataclysmic and non-cataclysmic production scenarios. Cataclysmic progenitors are predicted as relatively rare
explosive collisions between pairs of white dwarfs, or of old neutron stars \citep[e.g.][]{Totani, Kashiyama, Fuller}, 
or supramassive neutron stars collapsing into black holes \citep{Falcke}. Examples of non-cataclysmic progenitors are 
more common --- periodic flares or outbursts from rotating neutron stars like pulsars and magnetars
\citep[e.g.][]{Pen, Connor, Wasserman}. The explosion-based models suggest that FRBs are energetic, singular 
events while the outbursts-based models suggests that FRBs are 
less energetic and repeating. The examples here by no means exhaust the various models proposed in the literature.
However, despite being broadly classified as FRBs, each pulse 
appears to be unique in its own way (in terms of temporal pulse broadening, scintillation, frequency structure and polarization properties) 
hindering a consensus for their origin.  

Several studies have reported the potential of using FRBs as a 
cosmological probe to solve the ``missing baryons" problem \citep{McQuinn, nat_keane},
acquire an independent measure of the dark energy equation of state \citep{Zhou}, and study the intergalactic magnetic field \citep{Zheng, RaviSci}.
Since FRBs are highly energetic pulses of coherent emission of very short duration, their polarization properties could prove
vital to understanding the unknown generation and evolution of magnetic fields in the Universe. 
Faraday rotation is a birefringence effect in which the angle of the linearly polarized light of a radio wave is rotated under the
influence of a magnetised plasma. The amount of rotation undergone is quantified by the rotation measure (RM) and can be approximated to be,

\begin{equation}
\mathrm{RM} = 0.810 \int_0^D \! n_{e} \, B_{||} \cdot dl \, \, \mathrm{rad} \,\mathrm{m^{-2}},
\label{eq:rmorig}
\end{equation}

\noindent where $n_{e}$ is the electron density along the line-of-sight (LOS) in particles cm$^{-3}$, 
$B_{||}$ is the vector magnetic field parallel to the LOS in microgauss ($\upmu$G) and $dl$ is the elemental
vector towards the observer along the LOS. The sign of the RM quantity is such that a field oriented towards the observer is positive and away from the observer is negative.
An important advantage of Faraday rotation is that it is observable in any plasma irrespective of whether or not those atoms/molecules have magnetically susceptible energy levels.
This indicates that we can ignore interstellar extinction and thus probe magnetic fields out to cosmological distances.
The magnitudes of pulsar RMs range from very small values or order less than $1 \, \mathrm{rad}\,\mathrm{m}^{-2}$ 
\citep{Han, Noutsos, Schnitzeler} to $(-6.696\pm0.005) \times 10^4 \, \mathrm{rad}\,\mathrm{m}^{-2}$ for the magnetar J1745$-$2900, near Sagittarius A* at the centre of our Galaxy \citep{Eatough, Desvignes}.
There can be up to $\sim 10$ rad m$^{-2}$ (also depending on the time of day) of contribution to the RM from the Earth's ionosphere \citep{Kronbergbook}.
Recently, the repeating FRB 121102 (also referred to as `the repeater' in this paper) discovered at the Arecibo telescope in Puerto Rico was found to be 100\% linearly polarised with an average 
RM $= (+1.027 \pm 0.001) \times 10^{5}$ rad m$^{-2}$ \citep{Michilli}, which is the largest measured RM of an extragalactic source to date. RMs of this 
magnitude have only been observed near the vicinities of black holes with masses $> 10^{4} \, \mathrm{M}_{\odot}$, an example of which is the supermassive black hole Sagittarius A* at the centre of our Galaxy, 
with RM $= (-4.3 \pm 0.1) \times 10^{5}$ rad m$^{-2}$ \citep{BowerSGR, Marrone}.

Of the 28 published FRBs\footnote{All published FRBs and their measured properties are available on the FRB Catalogue http://www.frbcat.org} \citep{frbcat}
only 6 have polarization information \citep{Petroff_FRB,Masui,nat_keane,RaviSci,Emily,Michilli} 
due to the fact that only Stokes I is preserved in typical searches. It should be noted that the FRBs discovered at the UTMOST telescope also have only Stokes I recorded due to 
the antennas being only right circularly polarised. The total intensity of the signal is defined as $I \geqslant \sqrt{Q^{2} + U^{2} + V^{2}}$, where $Q$ and $U$ are 
the linearly polarised components of $I$, and $V$ is the circularly polarised component of $I$. The total linear polarization is defined as $L = \sqrt{Q^{2} + U^{2}}$.

An important feature of FRBs, if they are indeed cosmological, is that the observed total rotation measure (RM$_\mathrm{tot}$) would then be a combination of different contributions,

\begin{equation}
\begin{split}
\mathrm{RM}_\mathrm{tot}  &=  \mathrm{RM}_\mathrm{iono} + \mathrm{RM}_\mathrm{Gal} \, + \, \mathrm{RM}_\mathrm{IGM}  \\
& + \mathrm{RM}_\mathrm{host} + \mathrm{RM}_\mathrm{source}
\end{split}
\label{eq:rm_contributions}
\end{equation}

\noindent where $\mathrm{RM}_\mathrm{iono}$ is the RM due to the Earth's ionosphere \citep{Sotomayor-beltran}, $\mathrm{RM}_\mathrm{Gal}$ is the Galactic component assumed 
to typically vary with Galactic latitude and longitude  
and includes the local universe RM contributions, $\mathrm{RM}_\mathrm{IGM}$ 
is the contribution from the intergalactic medium (IGM) in the form of galaxies or filaments of 
cosmological large-scale structures along the LOS, $\mathrm{RM}_\mathrm{host}$ is the RM due to anomalous regions in a host galaxy and
$\mathrm{RM}_\mathrm{source}$ is the ``intrinsic'' component from magnetised plasma associated
with the progenitor source and its immediate environment. In the case of a source at cosmological distances, the RM along the LOS after accounting 
for the contributions from $\mathrm{RM}_\mathrm{Gal}$ and $\mathrm{RM}_\mathrm{iono}$,
will be reduced by a factor of $(1 + z)^2$ as,

\begin{equation}
\mathrm{RM}(z) = 0.810 \int_0^z \! \frac{n_{e}(z) \, B_{||}(z)} {(1+z)^2} \cdot \frac{dl}{dz} \, dz,
\label{eq:rm}
\end{equation}

\noindent due to the redshifting of the observed frequencies. Based on the RM and DM, the average magnetic field along the LOS can be estimated as,

\begin{equation}
\label{eq:Bll}
\langle B_{\parallel} \rangle \simeq 1.232 \, \bigg(\frac{\mathrm{RM}} {\mathrm{rad\,m^{-2}}} \bigg) \bigg(\frac{\mathrm{DM}} {\dm} \bigg)^{-1} \; \mathrm{\upmu}\mathrm{G}.
\end{equation}

It should be noted that the above equation does not include the redshift correction, which can be accounted for by replacing DM by DM/$(1+z)$ \citep{AkahoriFRB}.
The average magnetic field along the LOS can be affected by: 1) the presence of electron 
density inhomogeneities such as a hot nebula or HII region \citep{BannisterMadsen} which will dominate and cause the magnetic field to be
overestimated; and, 2) field reversals occurring along the sight line similar to the ones we see between the spiral arms of the 
Milky Way \citep{Han}. In both cases the individual components in Equation \ref{eq:rm_contributions} cannot be measured independently and therefore 
the large scale magnetic field along a given LOS cannot be accurately estimated. 

For FRB 140514 observed at Parkes,  
a significant circularly polarized ($V = 21 \pm 7\%$) component was observed and no RM could be measured \citep{Petroff_FRB}. 
For FRB 110523 observed at the GBT, the signal was significantly linearly polarized ($L = 44 \pm 3\%$, $V = 23 \pm 30\%$) with a large RM of 
$-186.1.1 \pm 1.4 \, \mathrm{rad}\,\mathrm{m}^{-2}$  \citep{Masui}. FRB 150807 discovered at Parkes, was found to be highly linearly polarization with a fraction of $L = 80 \pm 1\%$ 
and an RM of $12.0 \pm 7 \, \mathrm{rad}\,\mathrm{m}^{-2}$  \citep{RaviSci}. However, the measured RM of FRB 150807 is probably dominated by the same
contributions seen towards the pulsar J2241$-$5236 ($+13.3$ rad m$^{-2}$) located along a nearby sightline. \cite{RaviSci} estimate the RMs due
to magnetic fields in the region of detection (Galactic halo at high Galactic latitude) to be much smaller than that of the pulsar and therefore adopt the pulsar RM as the Galactic contribution for the given LOS. 
FRBs 150215 \citep{Emily} and 150418 \citep{nat_keane} were also found to be 
linearly polarized  $L = 43 \pm 5\%$ and $L = 8.5 \pm 1.5\%$, respectively but yielded RMs consistent with zero. More recently, as anticipated above, the repeating FRB 121102 was found to be
100\% linearly polarised with a large RM of $(+1.027\pm0.001)\times10^5\, \mathrm{rad}\,\mathrm{m}^{-2}$. No obvious trend has emerged from the small sample of FRB RMs (see Table \ref{tab:comparison}).
\cite{Bhandari} report the detections of 4 new FRBs discovered in the SUrvey for Pulsars and Extragalactic Radio Bursts \cite[SUPERB;][]{Superb1}. Two
of these were detected in real-time with full polarization information. 
Here we present the polarization analyses of these FRBs 151230 and 160102.

\section{Observations and Analyses} 

\subsection{Observing setup}
\label{sec:processing}

SUPERB uses the 21-cm multibeam receiver \citep{Staveley-smith} at the Parkes radio telescope along with the Berkeley Parkes Swinburne Recorder (BPSR)
backend instrument to obtain data. BPSR records 8-bit full polarization data from two orthogonal linear feeds per beam for each of the 13
beams of the multi-beam receiver over 400 MHz bandwidth from 1182 to 1582 MHz with 1024 channels at 64-$\upmu$s time resolution. This 8-bit data
is sent to a 120-s ring buffer and is processed in real-time using the \textsc{heimdall} single pulse search software. If \textsc{heimdall} identifies a candidate matching 
the criteria in Equation 1 of \cite{Superb1}, it saves the 8-bit data in the buffer within the time window $t_{0} - \Delta t \leq t \leq t_{0} + 2\Delta t$, where $t_{0}$ is the 
time at which the event occurred at the highest frequency of the observing band, $t$ is the time elapsed since the start of the observation and $\Delta t$ is the
dispersive delay across the whole band for the total observed DM.

FRBs 151230 and 160102 reported in \cite{Bhandari} were detected in beams 03 (inner hexagon) and 13 (outer hexagon) of the multibeam receiver 
at DMs of $960.4 \pm 0.5 \, \dm$ and $2596.1 \pm 0.3 \, \dm$, respectively. Data from all 13 beams of the receiver were analysed 
in detail and no detection was visible in any of the neighbouring beams above a signal-to-noise (S/N) threshold of 6. 
Hence we classify them as single-beam detections.

\begin{figure}
\includegraphics[width=0.48\textwidth]{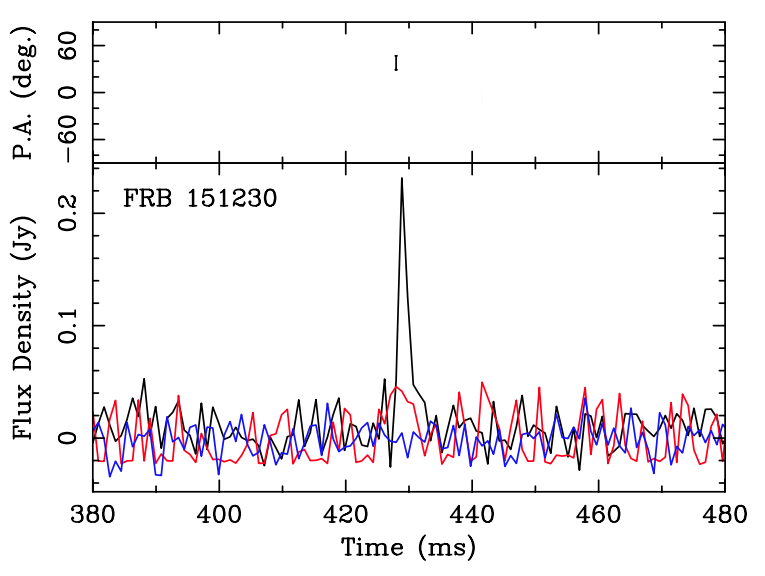}%
\par\medskip
\includegraphics[width=0.48\textwidth]{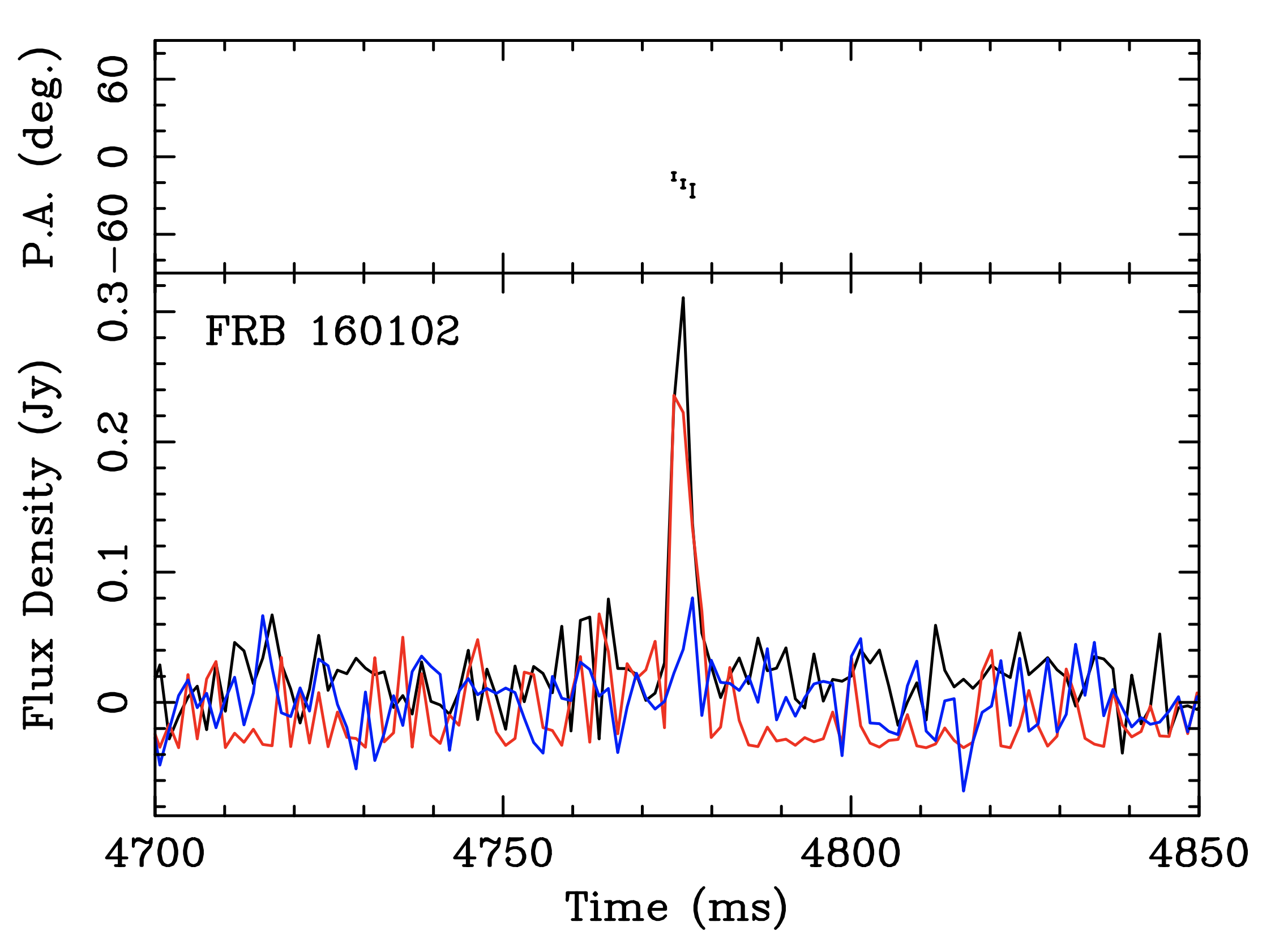}%
\caption{Polarization profiles of FRBs 151230 (top) and 160102 (bottom) after RM correction. Top panel: Polarization position angles of the electric field vectors 
across the on-pulse phase bins. Bottom panel: The total intensity (black), total linear polarization (red) and total circular polarization (blue) 
are shown as a function of time. FRB 151230 is seen to have linear and circular fractions of $L = 35 \pm 13\%$ and $V = 6 \pm 11\%$ assuming RM $= 0$ rad m$^{-2}$ (see text for details) and 
FRB 160102 is seen to have  linear and circular fractions of $L = 84 \pm 15\%$ and $V = 30 \pm 11\%$ after correcting for RM $=-220.6$ rad m$^{-2}$.
The data are not flux calibrated and the flux densities are in arbitrary units.}
\label{fig:poln}
\end{figure}

\subsection{Polarization calibration}
At Parkes, each of the 13 beams of the multi-beam receiver is equipped with two orthogonal linear feeds to measure the electric field vectors
$E_\mathrm{x}$ and $E_\mathrm{y}$, of the $X$ and $Y$ components of the incoming radiation. A calibration probe is placed at a $45^{\circ}$ angle to the
signal probes to enable the injection of a linearly polarized broad-band pulsed calibration signal.
We used standard calibration techniques from the \textsc{psrchive}\footnote{\url{http://psrchive.sourceforge.net/}} package. We adopt the ideal feed assumption 
(IFA) according to which the receptors are perfectly orthogonally polarized, and the reference source is 100\% linearly polarized and illuminates both receptors equally and in phase.
We do not adopt the Measurement Equation Template Matching (METM) model \citep{metm} as the uncertainty of the FRB position in the beam and the potential 
variation of polarimetric response with position in the beam make the precision of METM unnecessary.  
In keeping with the IAU/IEEE convention, for the Parkes 21-cm multibeam receiver we set the symmetry angle to $-\pi/2$ \citep{Willem_IAU}.
A calibration of the gain and phase of the receiver system was done by recording a pulsed calibration signal for 2 minutes after the detections. Calibration data
were recorded 45.4 and 1.2 hours after the detections of FRBs 151230 and 160102, respectively. We note that no changes were made to the observing set-up 
between the discovery and calibration observations of FRB 151230. 

\cite{RaviSci}, after extensive testing, conclude that the receptors in the central beam 01 of the multibeam receiver exhibit cross coupling at the 10\% level. 
Assuming the response of the other beams to be similar, \cite{RaviSci} expect the fractional error in measurement of their polarization for the FRB reported in their paper to be $< 10\%$.
The positional uncertainty of the FRB in the beam of detection might affect the degree of polarization. The attenuation of the polarization due to this uncertainty
has been empirically modelled and studied in \cite{RaviSci}. Tests to measure the RM and study the polarization profile of PSR J1644$-$4559, which is a good calibrator due
to its stable flux density, at various 
on- and off-axis beam positions for all the 13 beams were found to be consistent with the values and profiles in published literature \citep{Simon1644, Han}. Thus even in the case of 
the most extreme offset from the true position of FRBs, we do not expect the polarization measured to be significantly different from the intrinsic properties
except in the unlikely event that it is present in a sidelobe of the beams in the outer hexagon of the Parkes multibeam, in which case further investigation is required.

\section{Results}

We present the polarization profiles of FRBs 151230 and 160102 in Figure \ref{fig:poln}.
To determine the RM, the data were processed using both a brute force search method with \textsc{rmfit}\footnote{http://psrchive.sourceforge.net/manuals/rmfit/} described in \cite{Aidan}
and the RM synthesis method as described in \cite{Macquartetal2012}. 

\begin{figure*}
\includegraphics[width=5 in]{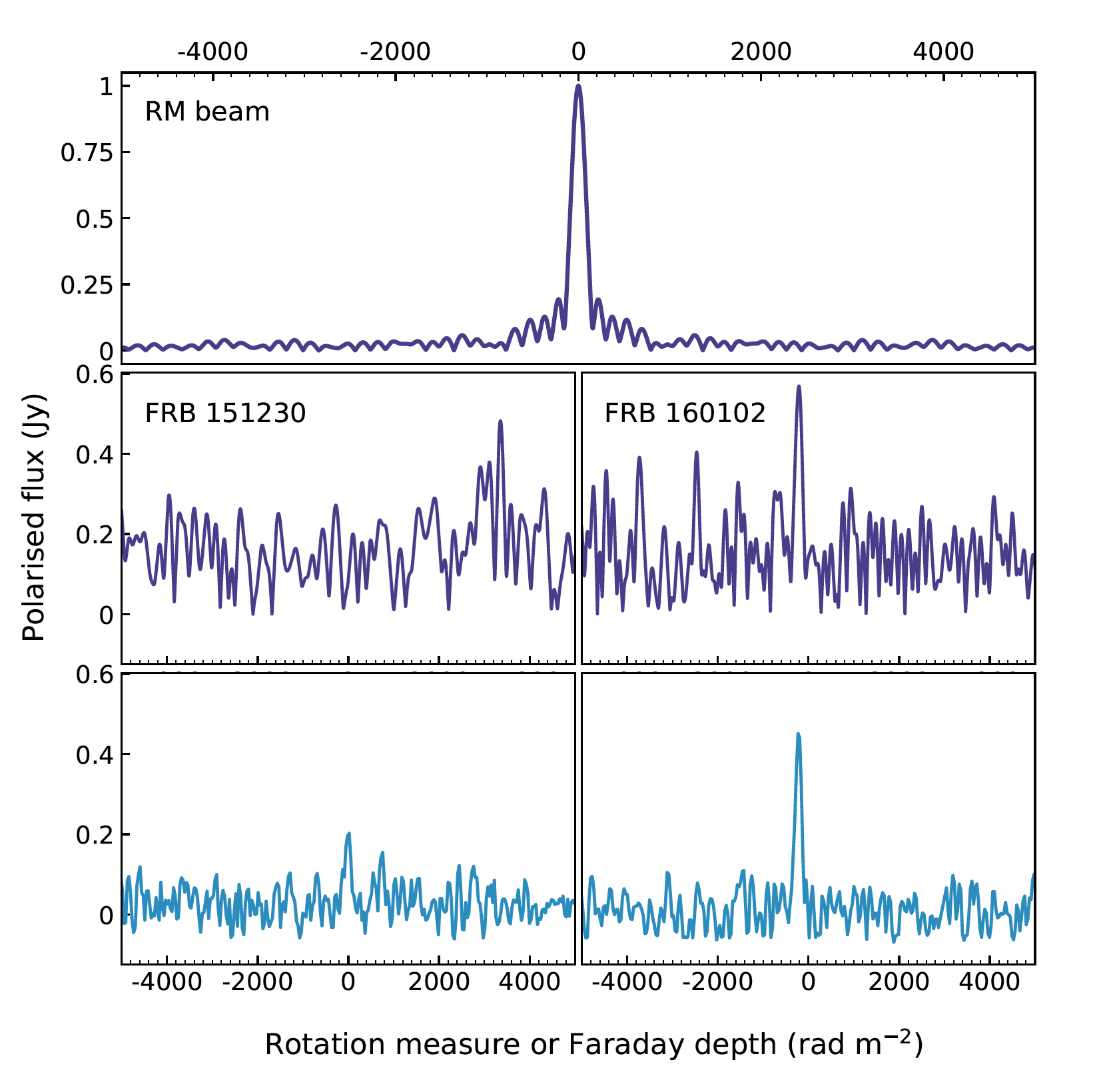}
\caption{The RM spectra (not deconvolved) of FRB\,151230 (middle, left) and FRB\,160102 (middle, right) from the RM synthesis show the amplitude of the polarized signal obtained by winding up the channelised ($Q$,$U$) vectors for a range of trial RM values as described in Section \ref{sec:JPrmsynthesis}. The top panel shows the RM ``beam'', which is the instrumental response to a polarised signal of unit amplitude at RM $=0\,\mathrm{rad\,{m}^{-2}}$.
Brute force searches for RM using \textsc{rmfit} for FRB 151230 (bottom, left) and FRB 160102 (bottom, right) were performed by looking for peaks in the
linearly polarised flux density.  A peak in the linearly polarised intensity is seen at 
RM $= -220.6\pm6.4\, \mathrm{rad}\,\mathrm{m}^{-2}$ for FRB 160102 and no significant RM was recovered for FRB 151230.
The polarization amplitude is in arbitrary units.}
\label{fig:rmsearch}
\end{figure*}

\begin{table*}
\centering
\caption{ Summary of the polarization properties measured in the FRB sample available so far. The second and third columns list the observed fractional linear and circular polarization, respectively.}
\label{tab:comparison}
\vspace*{5mm}
\centering
\begin{tabular}{c c c c c c}
\hline \\ [0.01ex] 
FRB Name &    $L$(\%) &   $V$(\%)  & Total RM (rad m$^{-2}$) & Galactic RM (rad m$^{-2}$) & Reference \\ [1ex] 
\hline \\ [0.01ex] 
110523 & 44 $\pm$ 3 & 23 $\pm$ 30 & $-186.1\pm 1.4$ & 18 $\pm$ 13 & \citet{Masui} \\
121102 & 100 & -- & $+ 1.026\pm0.001 \times 10^{5}$  & $-25 \pm 80$ & \cite{Michilli} \\
140514 & $<$10 & 21 $\pm$ 7 & -- & -- & \citet{Petroff_FRB}\\
150215 & 43 $\pm$ 5 & 3 $\pm$ 1 & $-9 < \mathrm{RM} < 12$ & -- & \citet{Emily} \\
150418 & 8.5 $\pm$ 1.5 & -- & 36 $\pm$ 52 & -- & \citet{nat_keane} \\
150807 & 80 $\pm$ 1 & -- & 12.0 $\pm$ 7 & 13.3 & \citet{RaviSci} \\
151230 & 35 $\pm$ 13 & 6 $\pm$ 11 & -- & -- & This work \\
160102 & 84 $\pm$ 15 & 30 $\pm$11 & $-220.6 \pm 6.4$ & 24.6 & This work \\ [1ex] 
\hline
\end{tabular}
\end{table*}

\begin{table*}
	\centering
	\caption{Derived properties of the two FRBs based on the results of RM synthesis. See Section \ref{sec:JPrmsynthesis} for more information.}
	\label{tab:RMsynthResults}
	\begin{tabular}{lccccc} 
		\hline\\ [0.01ex] 
		Event & RM (rad\,m$^{-2}$) & $\sigma_{\rm RM}$ & S/N & Polarization amplitude (Jy) & Reduced $\chi^2$ \\ [1ex] 
		\hline\\ [0.01ex] 
		FRB151230 & 3355 & 15 & 6 & 0.57 & 2.4\\
		FRB160102 & -210 & 13 & 6 & 0.48 & 1.4\\ [1ex]
		\hline
	\end{tabular}
\end{table*}

\subsection{Brute force search with RMFIT}
\label{sec:rmfit}

In the brute force search method with \textsc{rmfit}, we determine the RM by integrating in frequency for a range of trial RMs between
$\pm 1.2 \times 10^{5} \, \mathrm{rad}\,\mathrm{m}^{-2}$ in steps of $1 \, \mathrm{rad}\,\mathrm{m}^{-2}$ and search for a peak in
the total linearly polarised intensity $L = \sqrt{Q^{2} + U^{2}}$ (see Figure \ref{fig:rmsearch}). The searched RM range was determined by accounting for the maximum RM that we can measure within a single
channel after which the signal would be completely depolarized. The Stokes parameter profiles are produced for the upper and lower band profiles after correcting 
for the RM value corresponding to the peak in linearly polarized intensity in the previous step.
The best estimate of the RM is then obtained by taking the weighted mean position angle difference between the two bands with the weight inversely 
proportional to the square of the error in position angle difference for each pulse phase bin. 

The linear polarization never exceeded the S/N threshold of $3\sigma$ in the case of FRB 151230 and
no RM value could therefore be recovered. This FRB is seen to have average linear and circular polarization fractions of  
$L = 35 \pm 13\%$, $V = 6 \pm 11\%$ at RM $= 0$. We however recover an RM of $-220.6 \pm 6.4 \, \mathrm{rad}\,\mathrm{m}^{-2}$ for FRB 160102 as shown in 
Figure \ref{fig:rmsearch}. This FRB is seen to have average linear and circular polarization fractions of $L = 84 \pm 15\%$, $V = 30 \pm 11\%$ at the measured RM. 
The measured RM implies a LOS magnetic field strength of $B \gtrsim 0.1 \, \upmu$G, which is a lower limit due to possible LOS magnetic field reversals caused by intervening 
components like filaments or galaxies. 

\subsection{Rotation measure synthesis}
\label{sec:JPrmsynthesis}

The technique of RM synthesis \citep{Brentjens} was applied to the channelised $Q$ and $U$ measurements of FRBs 151230 and 160102.  
For each trial RM value, this algorithm calculates the amplitude of the summation of the channelised vectors ($Q$, $U$) after derotation by the appropriate 
angle for that frequency and RM value.  At trial RM values that do not correspond to the correct RM value, the ($Q$, $U$) vectors add incoherently, and the 
resultant polarization amplitude is consistent with the amplitude of the noise in the RM spectrum.  However, when the trial RM corresponds to the true RM of the 
signal, the derotated ($Q$, $U$) vectors sum coherently, and equal the amplitude of the polarized signal.
We used the implementation of this algorithm described in \citet{Macquartetal2012} to search for Faraday rotation of the polarized signals on grids with 
$1$ rad m$^{-2}$ resolution over Faraday depths spanning the range $\pm1.2 \times 10^{5}$ rad m$^{-2}$ to be consistent with Section \ref{sec:rmfit}.
RM synthesis includes the additional step of using a variation of \textsc{clean} to deconvolve the RM transfer function from the approximation 
to the Faraday dispersion function. This additional step differentiates RM synthesis from the brute force search implemented by \textsc{rmfit}. 

Table \ref{tab:RMsynthResults} lists the Faraday depths of the highest S/N polarised features detected and the resulting reduced $\chi^2$ value when a signal with the amplitude and RM obtained is 
applied to the data. In fact, it is important 
to test any signal against the hypothesis that it is a spurious noise spike.  One obvious test of any putative signal is to examine to 
what extent a signal of the derived amplitude and Faraday depth, when subtracted from the raw polarization data, results in a measurable decrease in the variance of the remaining 
signal. Ideally, the residual of a signal of the correct amplitude and Faraday depth subtracted from the data should be consistent with thermal noise. Thus, subtraction of the correct 
polarization model from the raw data should have a reduced $\chi^2$ value close to unity.  A signal whose subtraction from the raw data results in little or no reduction in the reduced 
$\chi^2$ is unlikely to be real.
Figure \ref{fig:rmsearch} shows the amplitude of the polarized signal as a function of Faraday depth for each of these two bursts before deconvolution.

For FRB 160102, the S/N of the polarized signal associated with a Faraday depth of $-210 \pm 13$ rad m$^{-2}$ reaches 6, and subtraction of this signal from the data 
yields a residual signal that is consistent with noise (with a reduced $\chi^2$ value of 1.4). This value of RM for FRB 160102, is consistent with the value obtained with \textsc{rmfit} in the previous section.
However, we are less confident of the peak in signal at RM $=3355 \pm 15$ rad m$^{-2}$ detected in FRB 151230 (see Table \ref{tab:RMsynthResults}).  
Although its S/N of 6 suggests that the signal is real at a $>98$\% confidence 
level\footnote{See Table 1 and associated text in \citet{Macquartetal2012} for a discussion of confidence as a function of S/N.}, 
various benchmark tests of RM synthesis codes against objects of known 
RMs show that events at S/N $\sim$ 6 - 7 are close to the threshold of believability for this algorithm \citep{Macquartetal2012, SchnitzelerLee}.  This is because the 
distribution of polarization amplitudes returned by the RM search is positive, definite and non-Gaussian, with a tail that extends to high S/N values; this is important in the 
present case, in which the search spans a range of $2.4 \times 10^{5}$ rad m$^{-2}$ in Faraday depth, with an RM beamwidth of approximately 120 rad m$^{-2}$. 
Moreover, the reduced $\chi^2$ after this polarized signal is applied to the data, is much larger at $2.4$ thereby making unlikely the possibility of it being real.  

Finally, in addition to the peak at 3355 rad m$^{-2}$ in the FRB 151230 spectrum, peaks at $\sim2500$ rad m$^{-2}$ and $\sim4000$ rad m$^{-2}$ 
in the FRB 160102 spectrum are also not robust to changes in the parameters of the search grid, indicating that they are unlikely to be genuine. The RM from FRB160102 however 
survives all of these tests giving us confidence that is a genuine astrophysical RM signature.

\begin{figure}
\includegraphics[width=3.3 in]{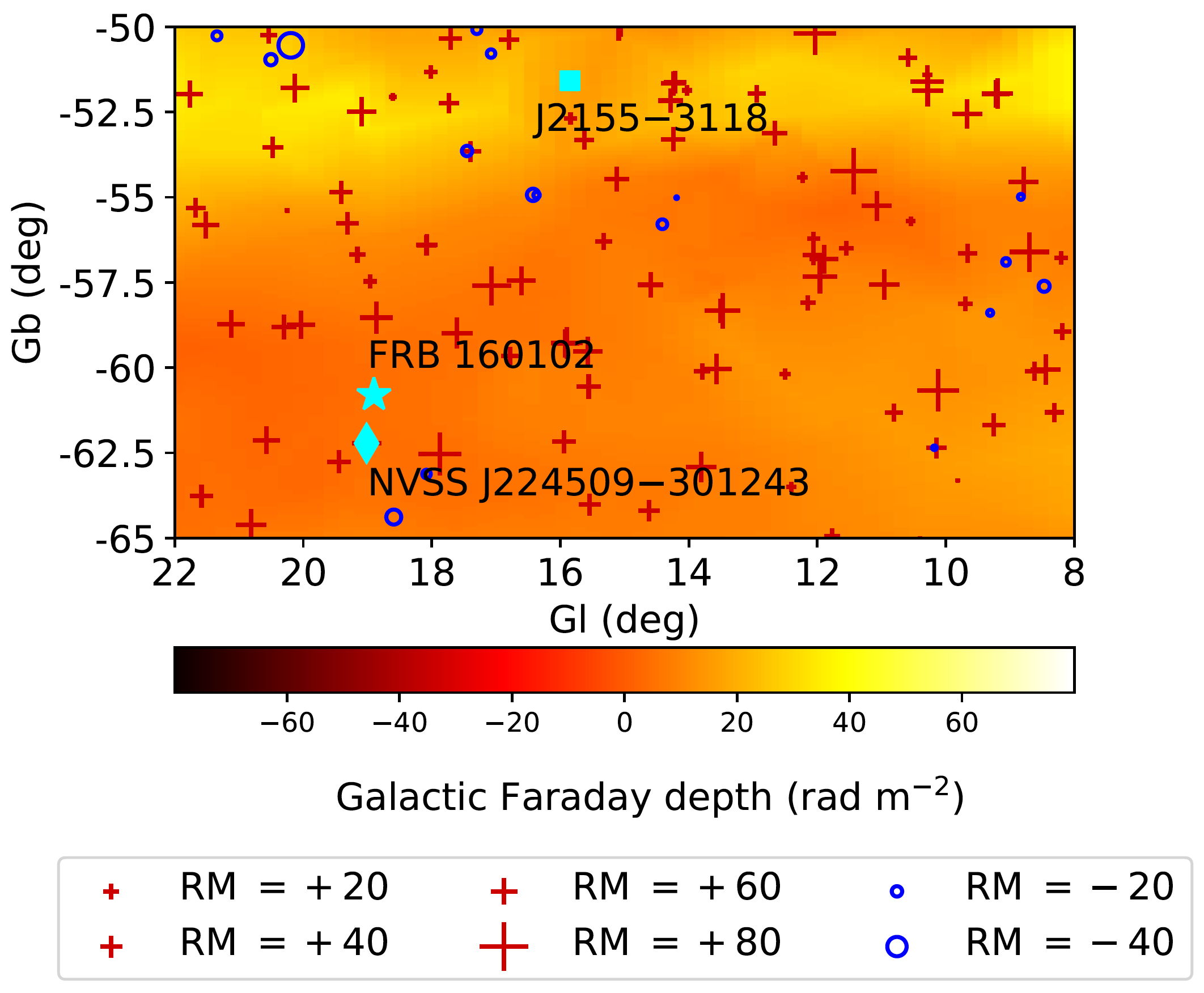}
\caption{Calibration of the Galactic foreground contribution to the observed RM for FRB 160102. The map represents the Galactic Faraday depth in rad m$^{-2}$ from \citet{Oppermann}. The negative and 
positive RMs from the NVSS catalog \citep{taylor} are represented by open circles and crosses, respectively. The sizes of the symbols are proportional to the RMs. 
The closest NVSS source and pulsar are represented by the diamond and square, respectively.}
\label{fig:foreground}
\end{figure}

\subsection{Galactic foreground and extragalactic contribution to rotation measure}
The RM contribution to FRB 160102 from the Galactic foreground was estimated in three different ways: using
nearby extragalactic polarised sources, RMs of pulsars in our Galaxy and from available RM maps (see Figure \ref{fig:foreground}).
In the first method, we identified the RM of the closest known NRAO VLA Sky Survey \citep[NVSS;][]{taylor} source. 
NVSS J224509$-$301243 is $\sim 2$ deg away from the position of the FRB, has an RM of $+28.7 \pm 6.3$ rad m$^{-2}$. 
In the second method, the closest known pulsar J2155$-$3118 
is $\sim 10$ deg away and has an RM of $+33.8\pm14.4$ rad m$^{-2}$ \citep{Han2018}. In the third method, the 
contribution from the smoothed Galactic foreground \cite{Oppermann} maps was estimated to be $+22 \pm 6$ rad m$^{-2}$. 
The foreground contributions from all three methods are seen to be consistent with an average of $+28 \pm 6$ rad m$^{-2}$.
The IGM can contribute $\lesssim \pm7$ rad m$^{-2}$ along a typical LOS out to the maximum estimated redshift of the FRB, $z \leq 2.1$ \citep{Oppermann}. Based on these values and the
RM obtained from \textsc{rmfit} we can 
estimate the $(\mathrm{RM}_\mathrm{source} + \mathrm{RM}_\mathrm{host})$ contribution from Equation \ref{eq:rm_contributions}, as 
$-256 \pm 9$ rad m$^{-2}$ not taking into account the redshifting of frequency.

\section{Discussion}
FRB 160102 is found to have an RM well in excess of what can be accounted for by both our Galaxy and the IGM. This is indicative of magnetisation in 
the immediate vicinity of the source or in the interstellar medium (ISM) of the host galaxy. Here we discuss different scenarios and possible progenitors for this FRB.

\subsection{Scenario 1: Origin in a nearby galaxy}
In this scenario, we assume that the bulk of the DM comes from the progenitor source and a host galaxy. 
FRB 160102's observed high DM and RM are indicative of either i) a rotation-powered scenario in which giant pulses are produced by young energetic pulsars with high spin-down energies 
\citep{Connor, Wasserman} or ii) a magnetically powered scenario in which radio emission accompanies magnetar hyperflares \citep{Popov, Lyubarsky, Maxim}.

\subsubsection{Pulsars associated with supernova remnants}
\label{sec:snr}

The influence of supernova remnants (SNRs) on FRBs produced by pulsars within them has been discussed in \cite{Connor, Maxim}, and \cite{Piro}. 
In this model, the energy released during a supernova explosion by ejecting some amount of mass $M_\mathrm{ej}$ outward with some velocity $v_\mathrm{ej}$
is,

\begin{equation}
E_\mathrm{SN} = \frac{1}{2} M_\mathrm{ej} v^{2}_\mathrm{ej}
\end{equation}

\noindent where $M_\mathrm{ej}$ is typically $1.4 \,\mathrm{M}_{\odot}$ for a Type Ia SNe and $\sim10 - 20 \, \mathrm{M}_{\odot}$ for a core-collapse SN, and 
$v_\mathrm{ej}$ is typically $10^{4}$ km s$^{-1}$ and 5000 km s$^{-1}$ for a Type Ia and core-collapse SNe, respectively \citep{Reynolds_snr}. The shock radius
scales as $R_\mathrm{s} = v_\mathrm{ej} t$. The outward expanding shock is not influenced by the ISM during the initial free expansion phase but heats it to
temperatures sufficient to ionise it. These electrons soon recombine due to the high density. However, as expansion continues interstellar gas accumulates
behind the blast shock as it propagates into the ISM, creating a reverse shock that propagates back into the ejecta material and re-ionizes it creating free electrons
that can account for the dispersion measure of an FRB.

The brightest pulse over the lifetime of the Crab pulsar would be visible out to $\sim300$ Mpc with current instruments, and similar pulses from other pulsars
within this distance could also produce FRBs. If every pulsar within a 100 Mpc distance produces $2 \times 10^{5}$ such pulses over its lifetime, it would be
sufficient to reconcile it with the estimated FRB rate \citep{Wasserman}. For a SNR at 200 Mpc 
the $\mathrm{DM_{source}}$ can be estimated as,

\begin{equation}
\begin{split}
\label{eq:DMs}
\mathrm{DM_{source}} & = \mathrm{DM_{tot} - DM_{host} - DM_{Galaxy}} \\
& \mathrm{- \,DM_{IGM}}\,\,\,\, \mathrm{pc \, cm^{-3}}
\end{split}
\end{equation}

\begin{figure}
\includegraphics[width=3.2 in]{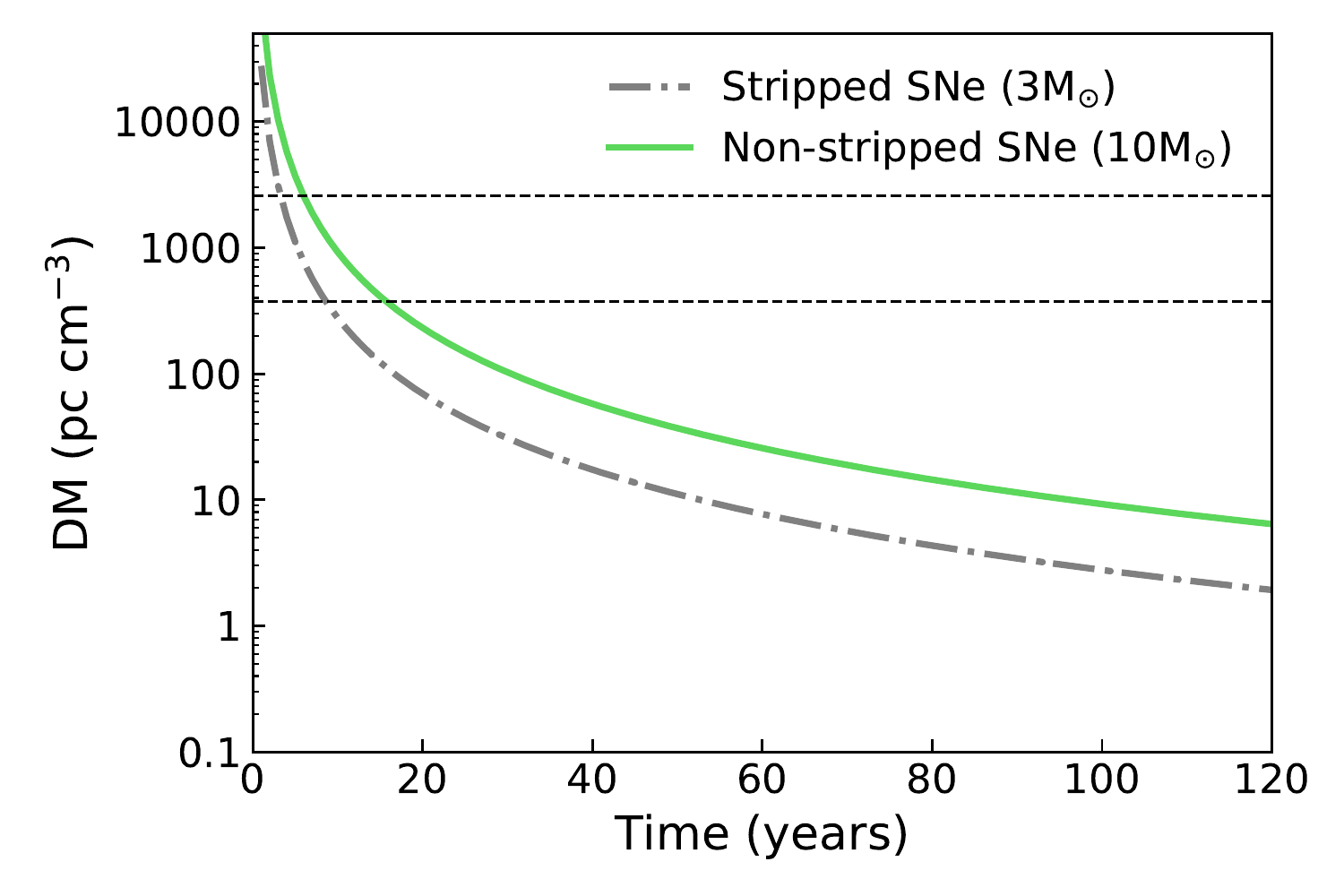}
\caption{DM as a function of age for stripped and non-stripped SNe following \citet{Piro}. Black dashed lines enclose the range of observed DMs for the known FRBs.}
\label{fig:DMageSN}
\end{figure}

\begin{figure}
\includegraphics[width=3.1 in]{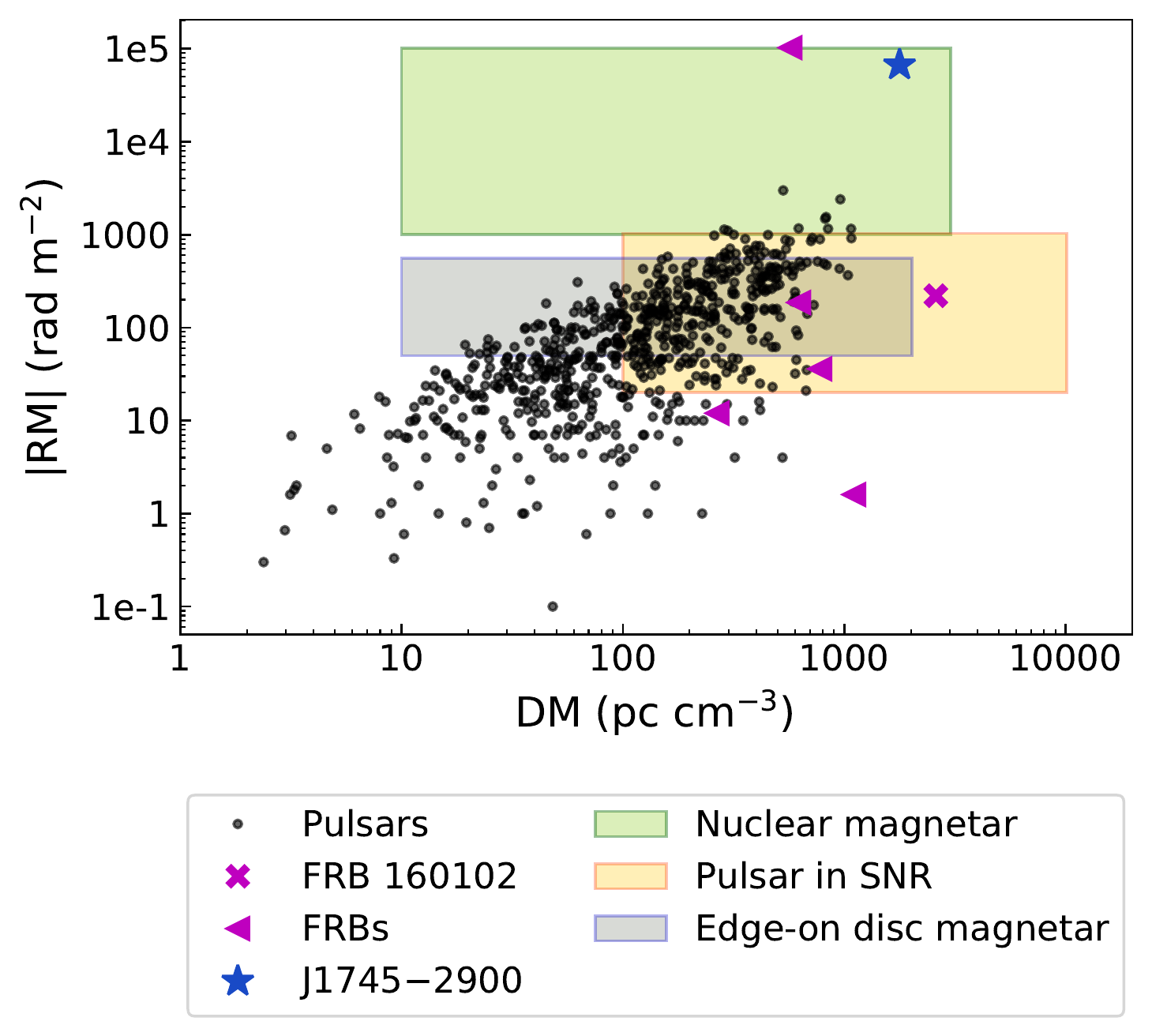}
\caption{Comparison between the RMs and DMs of pulsars and FRBs. The DM values for the FRBs are upper limits as they are the total observed values. 
The Galactic magnetar J1745$-$2900, has an unusual LOS which explains the high observed RM. The shaded regions enclose the RM and DM values expected from a pulsar associated with a SNR, 
a magnetar associated with a galactic centre and an edge-on disk magnetar.
FRB 160102 is seen to be similar to FRB 110523 in RM but with a much higher DM.}
\label{fig:DMsRMs}
\end{figure}

\noindent where the redshifting due to frequency is absorbed into the $\mathrm{DM_{source}}$ and $\mathrm{DM_{host}}$ terms. \cite{Bhandari} estimate the
contribution from the Milky Way for FRB 160102 to be $\sim 13 \, \dm$ and we assume the $\mathrm{DM_{host}}$ to be 100 $\dm$ for a median inclination angle of $< 70^{\circ}$
in a Milky Way type galaxy \citep{Cordes, Yao}.
Based on this we calculate the DM contribution from the SNR to be $\sim 2400 \, \dm$ for FRB 160102. Figure \ref{fig:DMageSN} shows the DM as a function of time for 
stripped (those SNe which have lost most of their initial hydrogen envelope, including most of their helium; 3 M$_{\odot}$) and non-stripped (10 M$_{\odot}$) 
SNe assuming an ejecta mass M$_\mathrm{ej} =$ 3 M$_{\odot}$ 
expanding with a typical velocity $v_\mathrm{ej} = 10^{4}$ km s$^{-1}$. A DM of $\sim 2400 \, \dm$ is expected at $\sim 3$ and $\sim6$ years after the explosion
for stripped and non-stripped SNe, respectively. \cite{Piro} show that free-free absorption in the SNR plays an important role in the propagation of the
FRB pulse and can dominate on timescales of $\sim 100-500$ years. This is backed up observationally by \cite{Bietenholz}. Their VLBI and 
VLA observations of the core-collapse supernova 1986J reveals a central, bright radio source which helps place constraints on the propagation of a radio pulse
through the SN ejecta. They conclude that the SN would have to be a few decades old before it is optically thin enough to a $\sim1$ GHz signal, by which time it would
not be possible for the ejecta to produce the observed high DM values seen in FRBs. If the SNR were to produce the observed DM after it becomes 
optically thin, it would require an unrealistically large portion of the ejecta to be ionized.
This implies that for FRB 160102 the majority of the DM is unlikely to originate in the vicinity of the progenitor, suggesting a cosmological origin. 

If FRBs are similar to giant pulses from the Crab pulsar with a similar birth period, and are emitted during the initial $10^{3}$ years of the progenitor's life time, the time interval between repeat pulse is tens
to hundreds of hours or longer. However if FRBs are rare and occur randomly during the source's lifetime (e.g. $\gtrsim 1$ Gyr), repeat pulses are seen over tens to hundreds of years or longer
\citep{Wasserman}. In such a scenario, it is difficult to place constraints on the repeat rates and classify an FRB as a one-off event or repeater. 

The Galactic large-scale magnetic fields obtained from Faraday rotation measurements of pulsars lie between 
$1.5 \, - \,2\, \upmu$G in magnitude. Fields can be as high as 11.2 $\upmu$G within the central 4 kpc Galactocentric radius \citep{Haverkorn}. More recently, \cite{Mao} have determined
the existence of coherent $\upmu$G magnetic fields in a galaxy at $z = 0.439$. The average LOS
magnetic field towards FRB 160102 without $z$ correction is $B > 0.1 \, \upmu$G which is well below the typical value from our Galaxy. \cite{Maxim} suggest that this could be due to the 
fact that the magnetic field giving rise to the RM is confined to the expanding shell. This is supported by \cite{Piro}, who show that the RM can be reasonably
explained by SNR magnetic fields and that the electrons producing the RM are likely different from the ones producing the DM, as the DM contribution is likely
small during the time when the radio pulse can propagate through the SNR environment. Therefore the model of a pulsar associated with a SNR in a nearby galaxy is ruled out for
FRB 160102 due to its inability to account for the large observed DM and the absence of a supernova within the error region of the FRB position. 

\subsubsection{Magnetars in galactic centres}
\label{sec:GCmagnetars}

The model of magnetars in galactic centres put forth by \cite{Pen} proposes that FRBs are produced by radio-loud magnetars located near the nuclei of galaxies within a few hundred Mpc
i.e. $z \ll 1$. The repeat timescales for hyperflares or similar events from magnetars are expected to be of the order of hundreds of years, except in the case where the magnetar is young 
and resides in an intensive star forming region \citep{Popov} in which 
case the timescales are shorter, thus providing an explanation for repeating FRBs \citep{Popov2018}.
The revelation that FRB 121102 originates in a dwarf galaxy, points to hydrogen-poor superluminous supernovae (SLSNe) and 
long gamma-ray bursts (LGRBs) as being possible progenitors of the source producing the radio pulses. These classes of transient events have been theorized to produce 
magnetars with high magnetic fields
of $\sim10^{14}$ G and millisecond spin periods \citep{Metzger, Nicholl}. 
The magnetar at the centre of our Galaxy, J1745$-$2900 has a measured DM of $1778\pm3 \, \dm$ and RM of $(-6.696\pm0.005) \times 10^4 \, \mathrm{rad}\,\mathrm{m}^{-2}$ \citep{Eatough}. 
The RM has since changed by 5.3\%, to RM $= (-6.340\pm0.023) \times 10^4 \, \mathrm{rad}\,\mathrm{m}^{-2}$ \citep{Desvignes} with only a marginal DM variation of 1\%
which is similar to what has been reported for the repeating FRB 121102 \citep{Michilli}.
Therefore in the case of a circumnuclear source in a nearby galaxy, one might expect RMs of $10^{3-5}$ rad m$^{-2}$ similar to the magnetar in our Galaxy \citep{Pen} which is inconsistent
with the observed RM of FRB 160102. 

Unlike FRBs whose scattering tails are typically $\sim$milliseconds broad, J1745$-$2900 is seen to be scattered to $\sim$second at $\sim$GHz frequencies. 
This scattering however is influenced by a screen closer to our Sun than the Galactic centre, as shown by VLBI measurements \citep{Spitler2014}. 
This implies that for an extragalactic observer, a typical LOS would only see scattering of the order of milliseconds. 
Circular polarization is significant in magnetar single pulses \citep{Kramer1810, Shannon, Eatough} and is not uncommon in other radio sources like pulsars. 
Single pulses from the Crab pulsar have been seen to be polarized with a high degree of linear and circular polarization
\citep{Slowikowska}. This circular polarization is consistent with both FRB 140514 \citep{Petroff_FRB} and FRB 160102.
Given that J1745$-$2900 is the only known radio-loud magnetar associated with a nuclear region, it is too soon to extrapolate and interpret the extent to 
which magnetars form preferentially in galactic centres. However the similarity in RMs between FRB 121102 and J1745$-$2900, and the significant difference in RMs
between FRB 160102 and J1745$-$2900 is suggestive of two progenitors for these FRBs.

\subsection{Scenario 2: Origin at cosmological distance}
\label{sec:explosion}

In this scenario, we assume that the bulk of the DM comes from the IGM with only a marginal contribution from the host galaxy and immediate vicinity of the progenitor source.

\subsubsection{Binary neutron star mergers}
\label{sec:bns}

All the models described in the previous sections imply repetitiveness. However, the vast majority of the FRBs have not yet shown repetition despite extensive 
monitoring in the radio. This could potentially be due to limited sensitivity, insufficient time on sky, the existence of more than one class of FRBs or a combination of these. 

Binary Neutron Star (BNS) mergers \citep{Totani, Yamasaki}  are cataclysmic events that justify the non-repeating nature of FRBs. 
The high end of the range for the BNS merger rate \cite[$\sim1 \times 10^{4}$ Gpc$^{-3}$ yr$^{-1}$;][]{Abadie} is close to the estimated FRB 
event rate \cite[$\sim2 \times 10^{4}$ Gpc$^{-3}$ yr$^{-1}$;][]{Thornton, Kulkarni}. In this model, the FRB is produced by radio emission at the time of the merger 
due to magneto-rotational energy production on a millisecond timescale. Since polarization is driven by the existence of magnetic fields, any radio emission from
BNS mergers can be expected to be polarized. \cite{Masui} favour supernova remnants and highly magnetised, dense star forming
regions for FRBs with high RM values (e.g. FRB 110523) and \cite{Yamasaki} favour much cleaner environments implying negligible magnetization in the circum-burst plasma,
like those expected around BNS mergers for FRBs with small RM
values (e.g. FRB 150807). In the case of FRB 160102, a BNS merger at a cosmological distance cannot account for the observed RM. Rather, a fraction of BNS mergers 
at cosmological distances leaving behind a stable remnant neutron star which may produce faint and repeating FRBs is preferred as this can account for both the observed 
DM and RM.

\subsubsection{Young stellar objects}
\label{sec:yso}

In this scenario for FRB 160102, we assume a DM of $\sim 2480$ pc cm$^{-3}$ to arise from the IGM after accounting for a Galactic contribution of $\sim 13$ pc cm$^{-3}$
and a host galaxy contribution of 100 $\dm$ for a median inclination angle of $< 70^{\circ}$ in a Milky Way type galaxy \citep{Cordes, Yao}. This places FRB 160102 at a redshift of $z \leq 2$ and
we calculate the RM at this redshift to be $\sim -2400$ rad m$^{-2}$ or larger in magnitude (see Equation \ref{eq:rm}). Such values of RM can be produced by models involving young stellar 
populations like magnetars \citep{Masui, Beloborodov}.

Similar to the SN model in Section \ref{sec:snr}, magnetars also emit an outflow of magnetized, relativistic winds resulting in a magnetar wind
nebula \citep[MWN;][]{Lyubarsky, Murase, Beloborodov}. An energetic pulse is generated as a result of the disconnection and reconnection of the magnetic field lines which 
propagates through the magnetar's relativistic wind and creates a relativistic forward shock upon interaction with the plasma. The millisecond duration radiation 
producing an FRB-like event would be due to coherent synchrotron maser emission from the shock front \citep{Popov2013, Lyubarsky, Beloborodov}. 
The ionized MWN produces the observed linear polarization and RM. This model predicts millisecond,
high energy TeV bursts to accompany the FRB pulse due to the synchrotron radiation at the forward relativistic shock front. However if the flare originates at distances greater than
a few to tens of Mpc, it is impossible for current $\gamma$-ray monitors to detect these bursts. The giant flare ($\sim 10^{42}$ erg s$^{-1}$), 
from the Galactic SGR 1806$-$20 on 27 December 2004 would only be detectable out to $\sim 30 - 40$ Mpc while the SWIFT telescope can detect similar flares out to
$\sim 70$ Mpc \citep{Hurley}. Observations by future observatories like the Cerenkov Telescope Array and VERITAS are expected
to be able to detect bursts from distances $\sim 100$ Mpc and can prove 
or falsify this model \citep{Lyubarsky, Murase, Popov2018}.

\begin{figure}
\includegraphics[width=3.2 in]{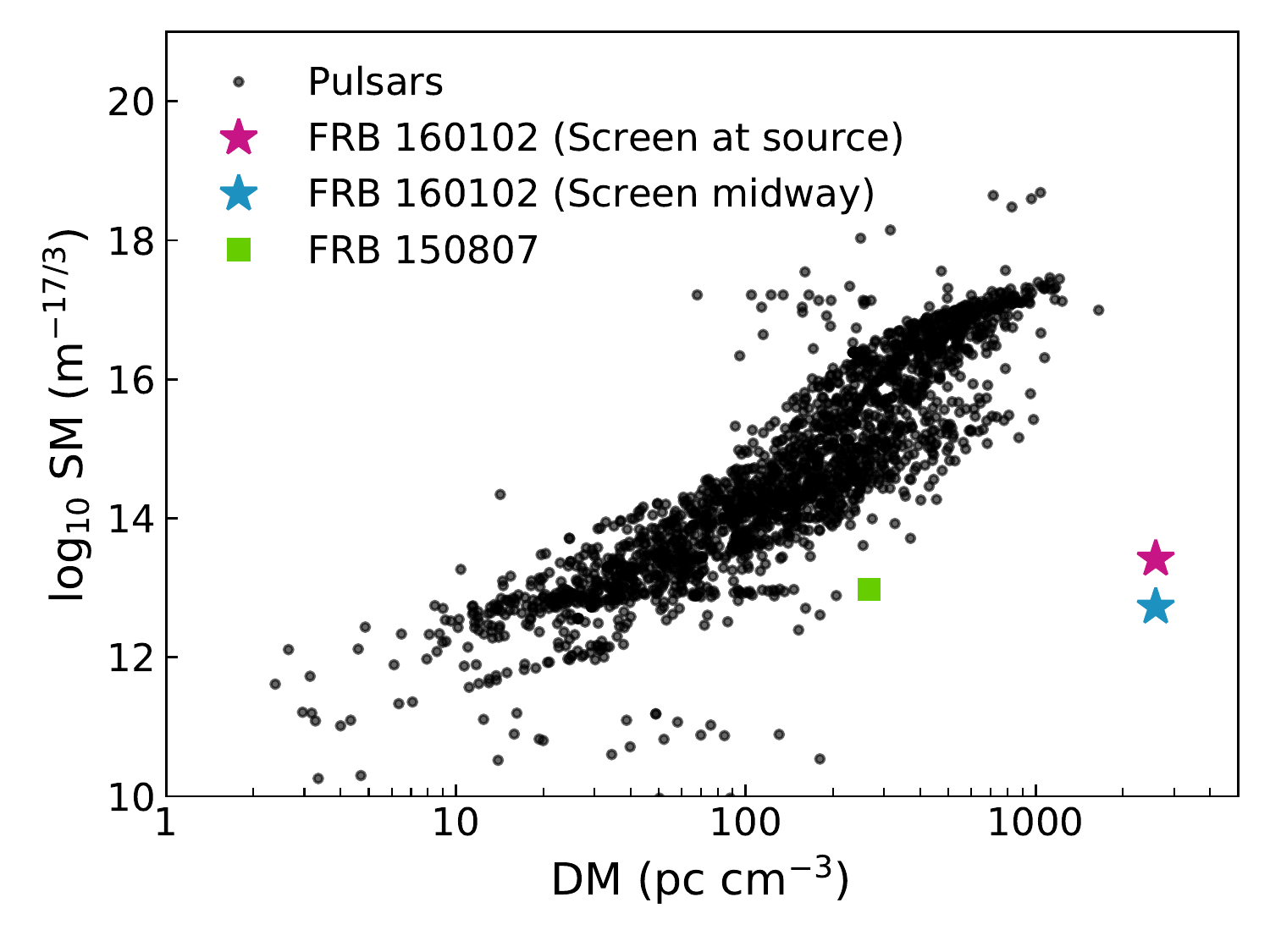}
\caption{Comparison of scattering measures (SMs) of pulsars and FRB 160102. }
\label{fig:SMsDMs}
\end{figure}

\subsection{Constraints on Scattering Measure}
Temporal broadening is caused by multipath propagation of radiation upon interaction with electron density inhomogeneities and turbulent plasma along the path of 
propagation thereby altering the observed pulse profile. The angles through which the frequencies are scattered relative to the radiation that travels along the LOS,
determine the time delay due to the divergence followed by the convergence back into the LOS. As the strength of the scattering increases so does the angle and time delay and is related to the LOS
integral of the plasma-density fluctuation power spectrum and is called the scattering measure (SM). FRBs which exhibit scattering, show temporal broadening consistent with
propagation through turbulent plasma \citep{Lorimer, Thornton, Champion, Macquart}. The contribution to the observed SM arises from the plasma fluctuations in the host galaxy 
of the FRB, the Milky Way and the IGM. We can rule out the possibility of Galactic contribution to the pulse broadening in the case of FRBs, as high Galactic latitude 
pulsars exhibit temporal broadening which
are orders of magnitude smaller \citep{Bhat, Krishnakumar}. According to \cite{Macquart} the scattering expected in the IGM for a given DM is much smaller 
than that due to the scattering seen in the Milky Way's ISM. \cite{Caleb_sim} in their Monte Carlo simulations of a cosmological population of FRBs show that they are unable to 
fit for the observed widths of the FRBs even after accounting for a scattering model for the IGM. They rule out the 
possibility of the scattering being LOS-dependent following \cite{Macquart}, given that the probability of interaction with the ISM of an intervening galaxy or an
intracluster medium along the LOS, is low. They conclude that the observed widths are likely to be dominated by a component intrinsic to the host galaxy or local environment. This conclusion is
backed by \cite{Cordes2016} who suggest that the inability of the IGM to contribute to the temporal scattering of the FRB pulse, implies that ionized regions in host 
galaxies must be responsible.

The pulse width of FRB 160102 is consistent with a temporally unresolved pulse due to a combination of uncorrected DM smearing in the individual frequency channels and interstellar 
scattering. Figure \ref{fig:SMsDMs} shows a comparison of the SMs for pulsars and two FRBs. A standard thin screen model is assumed at the source and mid-way 
between the source and observer for,

\begin{equation}
\begin{split}
\mathrm{SM} & = 2.73 \times 10^{17} (1+z_\mathrm{scr})^{17/6} (\nu/ 1\, \mathrm{GHz})^{11/3}  \\
& [\tau_\mathrm{s} (D_\mathrm{eff}/1 \, \mathrm{Gpc})^{-1}]^{5/6} \mathrm{m}^{-17/3}
\end{split}
\end{equation}

\noindent where $\nu$ is the observing frequency, z$_\mathrm{scr}$ is the screen redshift, and D$_\mathrm{eff} = (D_\mathrm{R}\, D_\mathrm{S})/(D_\mathrm{RS})$ where 
$D_\mathrm{R}$ is the observer-scattering screen distance,  $D_\mathrm{S}$ is the observer-source  distance, and $D_\mathrm{RS}$ is the source-screen distance. 
Given $\tau_\mathrm{s} = 3.95$ ms at 1 GHz \citep{Bhandari} and assuming the screen to be in the host galaxy ($z \sim 2$ for the total observed DM) we compute the SM to be 
$3 \times 10^{13}$ m$^{17/3}$. If the screen is mid-way between the source and observer,
($D_\mathrm{eff} = 2.7$ Gpc ; $z \sim 0.77$) we calculate the SM to be $5 \times 10^{12}$ m$^{17/3}$.  
From Figure \ref{fig:SMsDMs}, 
the value for FRB 160102 appears to be much lower than expected implying less turbulence in the IGM. A large sample of cosmological DMs and SMs would
prove invaluable in ascertaining the DM above which transients would be rendered undetectable.

\subsection{Effect of scattering on position angle}

It is well established from the observations of pulsars that scattering does not  only affect the total power of the shape of the total intensity pulse profile, but also its measured polarization properties 
\citep[e.g.][]{LiHan, KJ, Aris}. While all Stokes parameters are affected, including potentially the inferred degree of polarization, the impact of scattering is easily recognisable 
from a flattening of the position angle (PA) swing towards latter pulse phases \citep{LiHan}, which in the presence of so-called orthogonal polarization modes can also 
cause an apparent RM variation across the pulse phase \citep{Aris, Noutsos2015}. In the presence of scattering, similar effects should also be expected for 
FRBs. Due to the shortness of the burst duration, it is difficult to infer a swing of the PA in general, but in cases where PA variations may be seen, scattering would tend to 
wash out features, which could otherwise provide clues for the underlying origin of the FRBs. Even in cases where scattering is hardly detectable in total power, its impact on 
the PA may still be important and preventing a correct interpretation of the PA \citep[see][]{KJ, Aris}. The PAs for FRB 160102  in 
Figure \ref{fig:poln} indicate a negative slope with the suggestion of a flattening of the PA at the latest pulse phases. However, the PA uncertainties for the two FRBs are too large and the burst durations 
are too short to draw any firm conclusions, but we urge consideration of the impact of scattering when interpreting the polarization properties of FRBs in general.

\begin{figure}
\includegraphics[width=3.2 in]{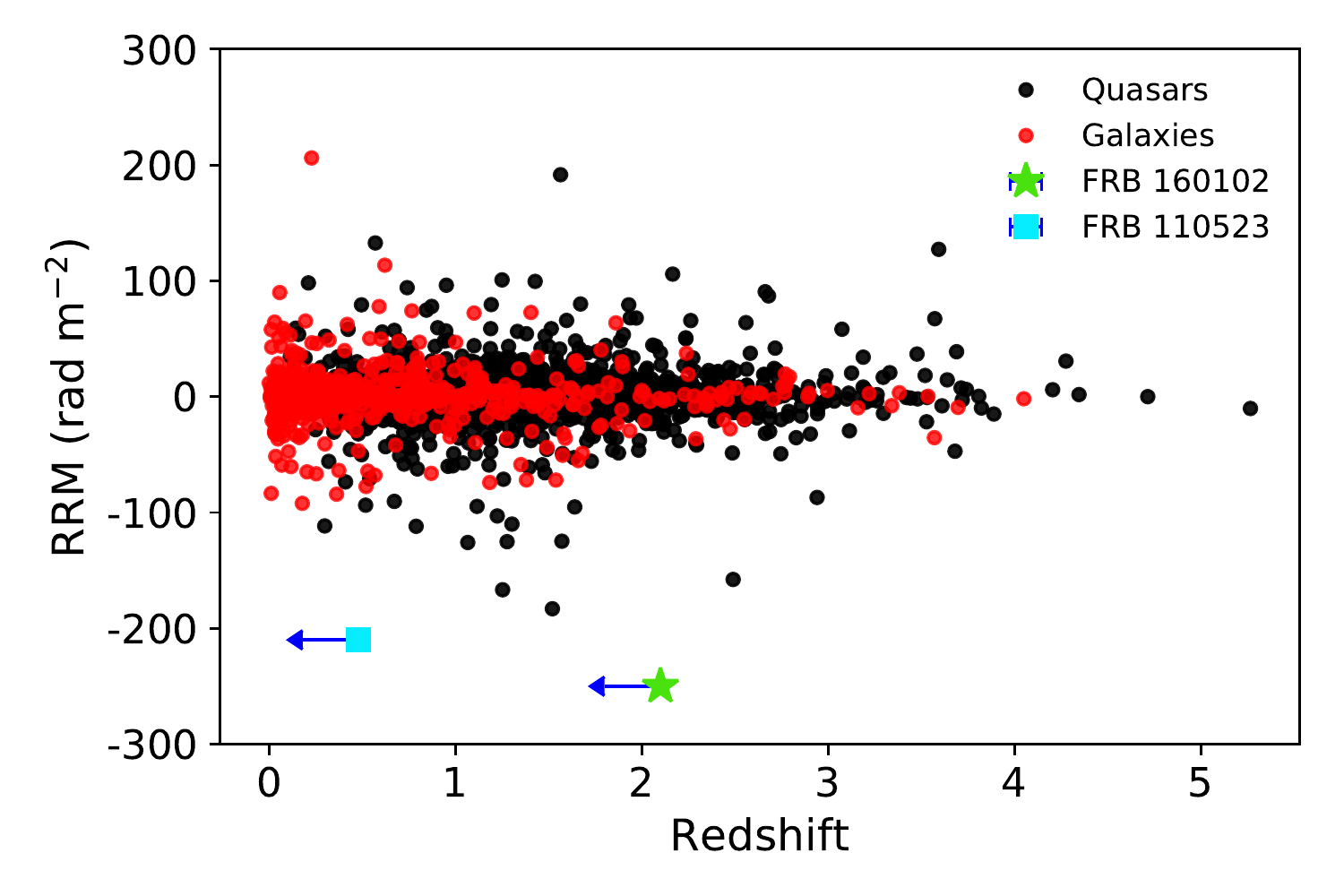}
\caption{Comparison of the residual rotation measures (RRMs) of galaxies and quasars with FRBs, as a function of redshift. The FRB redshifts are upper limits as they are inferred from their DMs, and would
decrease depending on the DM contribution to the host galaxy and local environment. FRB 121102 has been excluded from the plot for clarity.}
\label{fig:RRMs}
\end{figure}

\subsection{Extragalactic rotation measures}
Several studies have been used to probe the extragalactic magnetic fields by analysing the residual rotation measures (RRM; the difference between the observed RM and the Galactic RM contribution) 
as a function of redshift \citep{KronbergPerry, Reinhardt, Kronberg}. Though these studies report a mean RM of zero out to $z \sim 4$, the evolution of the RRM with redshift due to intervening
filaments/clouds and absorbers remains debatable. Simulations by \cite{Akahori} show that the RMs from the cosmic web would only be a few rad m$^{-2}$. \cite{Pshirkov} derived a statistically
robust Universe-wide magnetic field strength of $\lesssim0.65$ nG using the RMs of 3053 high latitude sources from the NRAO VLA Sky Survey catalogue. \cite{Mao} report 
measurements of coherent $\upmu$G magnetic fields in a lensed galaxy at $z = 0.439$ using broadband polarization data, similar in strength to galaxies in the local volume.

\citet{Hammond} report the largest RRM-redshift catalog comprising of 3650 high latitude sources. Figure \ref{fig:RRMs} shows the sources from their   
RRM-redshift catalog in comparison with FRB 160102. The FRB RM is inconsistent with the RMs of quasars and galaxies. 
In case of the FRB, a DM will accompany any RM and there may be multiple gas components which will contribute to both DM and RM with
varying proportions. Hence we cannot infer anything concrete from this comparison without an accurate localisation.

\subsection{Search for repeat pulses}

We have monitored fields of FRBs 151230 and 160102 over several months since their discovery to search for repeat pulses. We spent a total of 37.2 hours for FRB 151230 and 38.8 hours for FRB 160102 
with the Parkes multibeam receiver centred on the coordinates of the detection beam and 
detected no pulses with the real-time processing pipeline described in Section \ref{sec:processing}. We searched the data offline for pulses down to $\mathrm{S/N} > 5$ and still detected no 
pulses. Searches for repeats in these fields are ongoing, and strongly encouraged.

\section{Summary and Conclusion}

In this paper we present the properties of two FRBs reported in \cite{Bhandari}. FRBs 151230 and 160102 were found to have linear and circular polarization fractions of 
$L = 35 \pm 13\%, V =  6 \pm 11\%$ at RM $= 0$ rad m$^{-2}$ and  $L = 84 \pm 15\%, V =  30 \pm 11\%$ at RM $= -220.6$ rad m$^{-2}$, respectively. The average 
Galactic foreground contribution for the latter was estimated to be $+28 \pm 6$ rad m$^{-2}$ implying $\mathrm{{RM}_{source}} +  \mathrm{{RM}_{host}} = -256 \pm 9$ rad m$^{-2}$ ignoring the correction due to
redshifting of frequency and assuming $\mathrm{{RM}_{IGM}} = +7$ rad m$^{-2}$. 

Based on the observed RM we discuss possible progenitor scenarios for FRB 160102. In the case of a pulsar associated 
with a supernova remnant (see Section \ref{sec:snr}),
after accounting for the Galactic and a possible host contribution to the DM, the implied source DM of $\sim 2400$ pc cm$^{-3}$ is expected $\sim 3$ and $\sim 6$ years after the explosion for stripped and 
non-stripped SNe, respectively. The expected dominance of free-free absorption during these years would hinder the propagation of a radio wave. Therefore, the SNe would have to be at least a few decades
old before it is optically thin to a $\sim 1$ GHz signal by which time the ejecta would be too diffuse to contribute much to the large observed DM. 

The large DM and RM of this FRB along with the observed linear and circular fractional polarizations, can be explained by the model proposing that FRBs are from magnetars 
associated with galactic centres (see Section \ref{sec:GCmagnetars}). On the other hand, the very rare occurrence of known magnetars close to a galactic nucleus prevents any meaningful quantification of the likelihood of this hypothesis. 

We also consider explosion-based events such as BNS mergers at cosmological distances. While the BNS merger itself cannot quite explain the high RM value,
a stable NS remnant of the merger could potentially provide the observed high RM but would also imply repeatability (see Section \ref{sec:bns}).
Again, in the cosmological scenario, assuming the majority of the DM to arise from the IGM, we place FRB 160102 at $z \leq 2$.
The RM at source for this redshift is $\sim -2400$ rad m$^{-2}$ or larger in magnitude which is indicative of stellar models involving a young population such as magnetars (see Section \ref{sec:yso}).

The scattering measure for FRB 160102 based on the observed scattering at 1 GHz is seen to be much lower than what is expected thereby implying less turbulence in the IGM than usually assumed. 
We also observed a flattening of the polarization position angle swing for this FRB. However, a firm interpretation of that as due to the effect of scattering is hampered by the short duration of the burst 
and the uncertainty in measurement of the polarization position angle. We have compared the RM with the published extragalactic RMs of quasars and galaxies, but no secure association of FRB 160102
with a kind of known radio source could be established. 
It should be noted that multiple gas components along the LOS will alter the DM and RM. Thus no strong conclusions can be arrived at without accurate localisation of the FRB.

\section*{Acknowledgements}
MC would like to thank Bryan Gaensler, Stefan Oslowski and Tony Piro for useful discussion. 
MC and BWS acknowledge funding from the European Research Council (ERC) under the European Union's Horizon 2020 research and innovation programme (grant agreement No 694745).
EP acknowledges funding from the European Research Council under the European Union's Seventh Framework Programme (FP/2007-2013) / ERC Grant Agreement n. 617199.
The Parkes radio telescope is part of the Australia Telescope National Facility which is funded by the Commonwealth of Australia for operation as a
National Facility managed by CSIRO. Parts of this research were conducted by the Australian Research Council Centre of Excellence for All-sky Astrophysics 
(CAASTRO), through project number CE110001020 and the Laureate Fellowship FL150100148. This work was performed on the gSTAR national facility at Swinburne
University of Technology. gSTAR is funded by Swinburne and the Australian Government's Education Investment Fund.

\bibliographystyle{mnras}
\bibliography{poln}

\end{document}